\documentclass[twocolumn,plb,showpacs,superscriptaddress]{revtex4-1}%
\usepackage{amsfonts}
\usepackage{amsmath}
\usepackage{amssymb}
\usepackage{graphicx}%
\begin{document}

\title{Infinite Volume of Noncommutative Black Hole Wrapped by Finite Surface}
\author{Baocheng Zhang}
\affiliation{School of Mathematics and Physics, China University of Geosciences,
Wuhan 430074, China}
\author{Li You}
\affiliation{State Key Laboratory of Low Dimensional Quantum Physics, Department of Physics, Tsinghua University, Beijing 100084, China}

\begin{abstract}
The volume of a black hole under noncommutative spacetime background is found to be infinite,
in contradiction with the surface area of a black hole, or its Bekenstein-Hawking (BH) entropy,
which is well-known to be finite.
Our result rules out the possibility of interpreting the entropy of a black hole
by counting the number of modes wrapped inside its surface if the final evaporation
stage can be properly treated. It implies the statistical interpretation for the BH entropy
can be independent of the volume, provided spacetime is noncommutative.
The effect of radiation back reaction is found to
be small and doesn't influence the above conclusion.
\end{abstract}
\maketitle

\section{Introduction}

The interior of a black hole is not causally connected to its exterior. As a
result an external observer is prohibited from gaining any information about the
interior of a black hole, including the value of its volume.
The volume of a black hole was first investigated by Parikh \cite{mkp06}.
Subsequent studies invoked a slicing invariant definition with different time-like Killing vectors \cite{dg05,dm10,bl10,cgp11,bl13}.
Recently, Christodoulou and Rovelli (CR) suggested a different approach \cite{cr15},
which was followed by investigations in several spacetime backgrounds \cite{bj15,yco15,yco152}.
The CR volume $V_{\mathrm{CR}}$ is defined as the largest one bounded by
event horizon. Such a definition paints a different picture for the final stage
of a collapsed black hole, allowing it to possess
a very large interior instead of shrinking to a point.

The volume of a black hole is often discussed in connection
with the information loss paradox \cite{swh76}. The life of a
black hole is often considered to consist of two distinct segments:
formation followed by evaporation, assuming the latter starts only after the former
is completed. Such a hypothesis prompts a natural but seldomly asked question:
where is the information about the collapsed matter at the instant separating
the two segments? i.e., when formation is completed but evaporation is
yet to begin.

For Schwarzschild black hole, plausible answers consider information stored in
its interior, or distributed on its horizon as was initially postulated in the
\textquotedblleft quantum hairs\textquotedblright%
\ discussion \cite{kw89}. The existence of \textquotedblleft quantum
hairs\textquotedblright\ implies that information could reside on horizon
before evaporation. Recently, Hawking \textit{et al.}
show how to implant soft hairs on the horizon with soft degrees of freedom
proportional to the area of horizon in Planck units, based on physical
processes of the elegant mechanism using the soft graviton theorem \cite{hps16}.
The possibility that information can hide in the interior,
however, is not ruled out. To establish a firm answer, Parikh
suggested the illuminating idea \cite{mkp06} along the conversation between
Jacobson, Marolf, and Rovelli \cite{jmr05} to construct families of
spacetime whose horizon areas or surfaces are bounded but whose volumes wrapped inside
can be arbitrarily large. The entropies for such black holes must be
independent of their volumes. The amount of information associated with the
Bekenstein-Hawking (BH) entropies would have to be distributed over their horizons.

For stationary black holes in both three and four dimensions,
Parikh did not provide any preconceived constructions \cite{mkp06}.
Christodoulou and Rovelli show
$V_{\mathrm{CR}}\sim3\sqrt{3}\pi M^{2}v$ for a Schwarzschild black hole \cite{cr15},
with $M$ the (initial) mass and $v$ the advanced time.
This interesting result satisfies
the requirement of Parikh \cite{mkp06}, i.e., the volume becomes infinite for
$v\to\infty$. However, due to Hawking radiation \cite{swh74},
a Schwarzschild black hole evaporates and disappears at $v\sim M^{3}$.
Such an estimate for $v$
limits the volume for the (initial) black hole to a large but finite value $V_{\mathrm{CR}%
}\sim$ $M^{5}$.

Whether this volume is sufficient to house enough modes for explaining the
BH entropy statistically? A recent calculation \cite{zbc15}
implicates a negative answer, although it finds the
entropy calculated by counting modes housed in this volume is proportional
to the surface area. However, its treatment for the final evaporation
stage might be inadequate \cite{zbc15}.
An improved treatment would likely lead to a revised mass loss rate as
evaporation approaches the Planck scale, which will alter the estimated lifetime of a black hole.
This Letter presents our effort towards this by invoking
spacetime noncommutativity,
which is capable of treating the final evaporation stage without difficult singularities \cite{nss06,ans07}.

Adoption of spacetime noncommutativity leads to different black holes
\cite{ss031,ss032} and their associated thermodynamics (see \cite{pn09} for a
review and the references therein).
In noncommutative spacetime, the singularity in the interior of a black hole disappears,
and a remnant always arises irrespective \cite{acn87,coy15} after
black hole evaporation, which helps to remove the so-called Hawking paradox
of a diverging temperature as the black hole radius shrinks to zero.
We show such an approach overcomes the uncertainty of the earlier result \cite{zbc15}
by providing an improved description for the final evaporation stage,
with which we establish a firm answer to whether the interior of a noncommutative
black hole is large enough to explain the BH entropy.
We find a concrete example for a black hole with an infinite volume
but a finite horizon area.
Throughout this paper, we use units with $G=c=\hbar=k_{B}=1$.

\section{Noncommutative black hole}

We begin by briefly reviewing
the noncommutative Schwarzschild black hole and its
thermodynamics. The spacetime coordinate $x^{\mu}$ becomes
noncommuting \cite{nss06}
\begin{equation}
\left[  x^{\mu},x^{\upsilon}\right]  =i\theta \epsilon^{\mu\nu},
\label{non-com}%
\end{equation}
with the noncommutative parameter $\theta$ of dimension length squared.
It is a constant required by the Lorentz invariance and unitarity \cite{ss04}.
$\epsilon^{\mu\nu}$ is a real anti-symmetric tensor.
The commutation relation Eq. (\ref{non-com}) gives $\Delta
x^{\mu}\Delta x^{\nu}\geq\frac{1}{2}\theta$, the analogous Heisenberg uncertainty
relationship, which dictates the spacetime to be \textquotedblleft pointless\textquotedblright,
free from gravitational singularity.

The direct application of noncommutative coordinates to black holes is
inconvenient, so in this paper, we adopt the idea assumed in Ref. \cite{nss06} where
the spatial noncommutative effect is attributed to the modified energy-momentum
tensor as a source while the Einstein tensor is not changed.

In flat spacetime noncommutativity eliminates point-like structures in favor
of smeared distributions \cite{ss031,ss032}. When applied to spacetime in
gravity, one can simply make corresponding substitutions.
Instead of the Dirac $\delta$-function $\rho_{\theta}\left(
r\right)  =M\delta(r)$ usually employed for a point mass $M$ at origin in
commutative spacetime, the smearing leads to a Gaussian distribution
\begin{equation}
\rho_{\theta}(r) =\frac{M}{(4\pi\theta)^{\frac{3}{2}}}e^{-\frac{r^{2}}%
{4\theta}}, \label{nmd}%
\end{equation}
of a width $\sim\sqrt{\theta}$, with which the energy-momentum tensor was
identified for a self-gravitating droplet of anisotropic fluid \cite{nss06}.
Solving the Einstein equation gives
\begin{align}
ds^{2}  =& -\left[  1-\frac{4M}{r\sqrt{\pi}}\gamma\left(  \frac{3}{2}%
,\frac{r^{2}}{4\theta}\right)  \right]  dt^{2}\nonumber\\
&  +\left[ 1-\frac{4M}{r\sqrt{\pi}}\gamma\left(  \frac{3}{2},\frac{r^{2}%
}{4\theta}\right)  \right]  ^{-1}dr^{2}+r^{2}d\Omega^{2}, \label{nsc}%
\end{align}
with the lower incomplete gamma function
$\gamma(\nu,x) =\int_{0}^{x}t^{\nu-1}e^{-t}dt$,
which approaches $\sqrt{\pi}/2$ as $r\to\infty$.
For $\theta\rightarrow0$, $\gamma(\nu,x)$ reduces to the usual
$\Gamma(\nu)$-function and the noncommutative metric Eq. (\ref{nsc}) becomes
the commutative Schwarzschild metric.

The condition of $g_{tt}(r_{h})=0$ gives the event horizon
\begin{equation}
r_{h}=\frac{4M}{\sqrt{\pi}}\gamma\left(  \frac{3}{2},\frac{r_{h}^{2}}{4\theta
}\right)  \equiv\frac{4M}{\sqrt{\pi}}\gamma_{h}, \label{rh}%
\end{equation}
which takes the minimum $r_{h}^{(\min)}=r_{0}%
\simeq3.0\sqrt{\theta}$ at $M_{0}\simeq1.9\sqrt{\theta}$ determined by
${dM}/dr_{h}=0$ where the two horizons decay to one. In particular,
no horizon exists below $M_{0}$, which is not concerned in our paper.

\begin{figure}[ptb]
\centering
\includegraphics[width=2.75in]{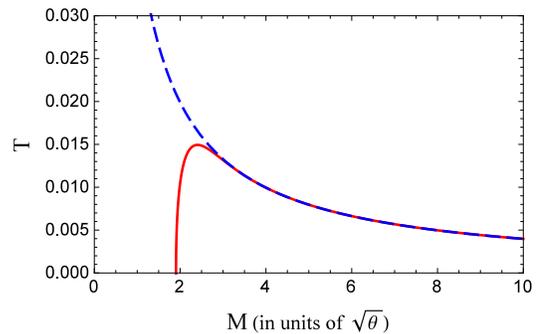}
\caption{(Color online) The temperatures $T$ for an evaporating Schwarzschild black hole
of (initial) mass $M$.
The red-solid (blue-dashed) line refers to
noncommutative (commutative) spacetime.  }%
\label{fig1}%
\end{figure}

The temperature is obtained for the static noncommutative metric Eq. (\ref{nsc}),
\begin{equation}
T_{h}=\frac{1}{4\pi}\left.  \frac{dg_{tt}}{dr}\right\vert _{r=r_{h}}=\frac
{1}{4\pi r_{h}}\left(1-\frac{r_{h}^{3}}{4\theta^{\frac{3}{2}} \gamma_{h}}
{e^{-\frac{r_{h}^{2}}{4\theta}}}\right),
\label{temp}%
\end{equation}
which reaches its maximum at $M\simeq2.4\sqrt{\theta}$ and decreases
to zero at $M=M_{0}$ as shown by the red-solid line in Fig. \ref{fig1}.
When $\theta\rightarrow0$, $T_{h}$ reduces to
$T_{H}={1}/({8\pi M})$, as for commutative spacetime denoted by the blue-dashed line.
From the first law of black hole thermodynamics $TdS=dM$, we find the entropy
\begin{equation}
S_{h}=\int \frac{dM}{T}\simeq
4\pi M^{2}\left(  1-\frac{4M}{\sqrt
{\pi\theta}}e^{-\frac{M^{2}}{\theta}}\right),
\label{ncbe}%
\end{equation}
up to the order of $e^{-{\frac{M^{2}}{\theta}}}/{\sqrt{\theta}}$.

\section{Volume and Entropy}

It was shown \cite{cr15,cl16} that the definition of CR volume can be applied to
Schwarzschild black holes, but also applied to any other spherically symmetric
spacetime. Therefore, as described in last section, the interior of
noncommutative black holes (\ref{nsc}) can define the CR volume. For this, one
has to rewrite the metric (\ref{nsc}) in terms of ingoing
Eddington-Finkelstein coordinates,%
\begin{equation}
ds^{2}=-f\left(  r\right)  dv^{2}+2dvdr+r^{2}d\varphi^{2}+r^{2}\sin^{2}\varphi
d\phi^{2}, \label{ef}%
\end{equation}
with $f(r)=$ $1-\frac{4M}{r\sqrt{\pi}}\gamma\left( \frac{3}{2},\frac{r^{2}%
}{4\theta}\right)$, and the advanced time $v=t+r^{\ast}$ for
$r^{\ast}\equiv\int^{r}\frac{dr^{\prime}}{f(r^{\prime})}$.
As shown before \cite{cr15,bj15}, the CR volume is
mostly related to the region which is not causally connected with matter that
has fallen far into the black hole, and so the contribution for the volume is
given by the integral%
\begin{equation}
V_{\mathrm{NCR}}=\int^{v}\max\left[F(r)\,\right]
dvd\varphi d\phi,
\end{equation}
where the integral will be dominated by its upper limit $v$ while the lower
integration limit is irrelevant, as pointed out in Ref. \cite{bj15}. Then we
calculate the maximal value of the function by setting $\frac{dF}{dr}=0$ for the
integrand $F(r)=r^{2}\sqrt{\frac{4M}%
{r\sqrt{\pi}}\gamma\left(  \frac{3}{2},\frac{r^{2}}{4\theta}\right)  -1}$,
and the maximum is found to occur at
\begin{equation}
r\simeq r_{n}\left(1-\frac{r_{n}}{\sqrt{\pi\theta}}e^{-{\frac{r_{n}^{2}%
}{4\theta}}}\right), \label{ncrm}%
\end{equation}
with $r_{n}={3}M/{2}$. Carrying out the integration, we find
\begin{equation}
V_{\mathrm{NCR}}\simeq3\sqrt{3}\pi M^{2}v\left(  1-\frac{2r_{n}}{\sqrt
{\pi\theta}}e^{-{\frac{r_{n}^{2}}{4\theta}}}\right),
\label{ncrv}%
\end{equation}
which is explicitly modified by $\theta$, apart from
 implicit $\theta$-dependence, e.g., in $v$.

Next we discuss the entropy associated with $V_{\mathrm{NCR}}$.
We change the metric Eq. (\ref{ef}) into the form
\begin{equation}
ds^{2}=-dT^{2}-\left[f(r)\dot{v}^{2}-2\dot{v}\dot{r}\right]  d\lambda^{2}
+r^{2}d\Omega^{2},
\end{equation}
with the transformation $dv=\frac{-1}{\sqrt{-f}}dT+d\lambda$ and
$dr=\sqrt{-f}dT$. Since the volume refers to the late time $v$ at
$r\simeq r_{n}\left(1-\frac{r_{n}}{\sqrt{\pi\theta}}e^{-{\frac{r_{n}^{2}%
}{4\theta}}}\right)$, we take the constant-$T$ hypersurface to
count the number of quantum field modes that can be housed in $V_{\mathrm{NCR}}$.
With suitable modifications consistent with the uncertainty
relationship \cite{skr01,cmt02,lx02}, the earlier method \cite{zbc15}
remains applicable to more general cases of quantum gravity effects,
including spacetime noncommutativity being discussed here.
The commutation between conjugate
position $x_{i}$ and momenta $p_{j}$ is unchanged, or $[x_{i}%
,p_{j}]=i\delta_{ij}$ \cite{ss031,ss032}, i.e., the uncertainty relation $\Delta
x_{i}\Delta p_{i}\sim2\pi$ retains. Now the phase space is
labeled by $\{\lambda,\varphi,\phi,p_{\lambda},p_{\varphi},p_{\phi
}\}$, and the volume element takes the form $d\lambda d\varphi
d\phi dp_{\lambda}dp_{\varphi}dp_{\phi}/(2\pi)^{3}$.

Compared with the commutative spacetime discussed earlier \cite{zbc15},
the only change concerns the factor $f(r)$ when computing noncommutative entropy.
We find the same form \cite{zbc15} for,
\begin{equation}
S_{\mathrm{NCR}}=\frac{\pi^{2}V_{\mathrm{NCR}}}{45\beta_{h}^{3}},
\label{ncrve}%
\end{equation}
with both $V_{\mathrm{NCR}}$ and $\beta_{h}$ modified by noncommutativity.
To study $S_{\mathrm{NCR}}$ in more detail, we need to specify the time $v$.
As shown clearly by the red solid line in Fig. \ref{fig1},
evaporation of a noncommutative black hole involves two stages;
before the maximal
temperature is reached, it is similar to what happens in commutative
Schwarzschild black hole; after that, the temperature decreases to zero
and a cold remnant of mass $M_{0}\simeq1.9\sqrt{\theta}$ is left behind.

First, we investigate the influence of back reaction.
As well-known, the dynamic evolution for a spherically symmetric black hole due to Hawking
radiation can be described by the Vaidya metric \cite{pv51}. With this metric,
the CR volume was already calculated and the change was found to be insignificant \cite{yco152,cl16},
where the estimation for advanced
time $v$ was made by using the Stefan-Boltzmann law (see Eq. (\ref{sbl}) below)
since the black hole was considered with the mass greater than the Planck mass \cite{sm95}.
However, the associated entropy is yet to be investigated for this case.

\begin{figure}[ptb]
\centering
\includegraphics[width=2.75in]{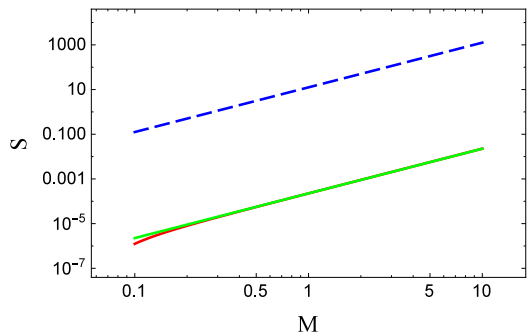}
\caption{(Color online) Entropies $S$
for a commutative Schwarzschild black hole (of initial mass $M$).
The red (green) solid line denotes the entropy
associated with the volume when back reaction is included (excluded).
The blue dashed line denotes the noncommutative BH entropy $S_{h}$ of Eq. (\ref{ncbe}).
}%
\label{fig2}%
\end{figure}

Transforming the mass into $M'(v)=\Theta(v)(M^{3}-3Bv)^{\frac{1}{3}}$ with
$\Theta(v)$ the Heaviside step function and the parameter $B\sim10^{-3}$
related to the back reaction \cite{sm95}, the CR volume
can be reexpressed as \cite{yco152,cl16},%
\[
V_{\mathrm{CR}}^{\prime}\simeq3\sqrt{3}\pi M^{2}v\left(  1-\frac{9B}{2M^{2}%
}\right)  ,
\]
with which we can compute the entropy associated with CR volume including the
effect of back reaction using Eq. (\ref{ncrve}).
Figure \ref{fig2} illustrates clearly that the entropy
associated with the volume is insufficient for a statistical
interpretation of the Bekenstein-Hawking entropy, even
when the back reaction is included. This interesting result
refutes against the possibility of
balancing black hole information loss by a huge but finite volume \cite{cr15,cl16,ao16}.
It is also seen from Fig. \ref{fig2} quantitatively
that the influence of back reaction is small for
our purpose. In what follows we will not include modifications
from the back reaction in our calculations.

According to Ref. \cite{mkp07}, in the first evaporation
stage, the mass loss rate for a black hole is given by the Stefan-Boltzmann law
\begin{equation}
\frac{dM}{dv}=-\frac{1}{\gamma M^{2}},\ \ \gamma>0. \label{sbl}%
\end{equation}
The specific value for the constant $\gamma$ does not influence the
present study. Integrating from an initial black hole mass $M$, one finds
$v\sim\gamma(  M^{3}-M_{f}^{3})  \sim\gamma M^{3}$, with
$M_{f}$ ($\ll$ $M$) being the critical mass where Eq. (\ref{sbl}) becomes invalid.
Assuming the first evaporation stage dominates $v$, omitting the second evaporation stage,
and with the definitions for $M$ and $\beta_{h}=T_{h}^{-1}$, we find
\begin{equation}
S_{\mathrm{NCR}}\sim\frac{\sqrt{3}\gamma M^{2}}{7680}\left(  1-\frac{3M e^{-\frac{9M^{2}}{16\theta}}}%
{\sqrt{\pi\theta}}\right)  \left(1-\frac
{2M e^{-{\frac{M^{2}}{\theta}}}}{\sqrt{\pi\theta}}\right)^{3}.
\nonumber
\end{equation}

\begin{figure}[ptb]
\centering
\includegraphics[width=2.75in]{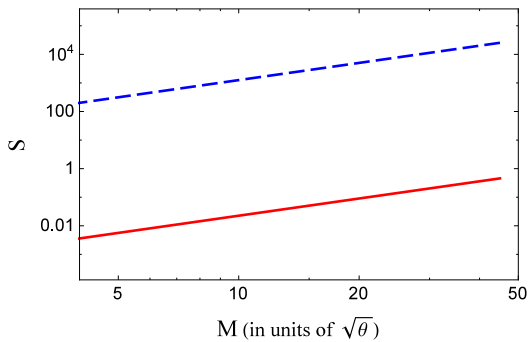}
\caption{(Color online) The entropy $S$ associated with the volume
of a noncommutative black hole in red solid line
with estimated $v$ limited to the first evaporation stage.
The blue dashed line denotes the noncommutative BH entropy $S_{h}$. }%
\label{fig3}%
\end{figure}

The surface area for a noncommutative black hole can be expressed approximately
as $A_{h}=16\pi M^{2}\left(  1-\frac{2M}{\sqrt{\pi\theta}}e^{-\frac{M^{2}%
}{\theta}}\right)^{2}$. Thus the entropy $S_{\mathrm{NCR}}$ 
associated with the noncommutative volume is proportional
to the horizon area of a noncommutative black hole.
We can rewrite it as $S_{\mathrm{NCR}}\sim
\frac{\sqrt{3}\gamma\varepsilon\left(  \theta\right)  }{122880\pi}A_{h}$,
where $\varepsilon\left(  \theta\right)  =\left(  1-\frac{3M}{\sqrt{\pi\theta
}}e^{-\frac{9M^{2}}{16\theta}}\right)  \left(  1-\frac{2M}{\sqrt{\pi\theta}%
}e^{-{\frac{M^{2}}{\theta}}}\right)$ approaches unity if
$v$ is approximated by the first evaporation stage.
Figure \ref{fig3} compares $S_{\mathrm{NCR}}$ with $S_{h}$.
The entropy associated with the noncommutative volume remains clearly
insufficient for a statistical interpretation of the BH entropy
if $v$ only accounts for the first evaporation stage.

\begin{figure}[ptb]
\centering
\includegraphics[width=2.75in]{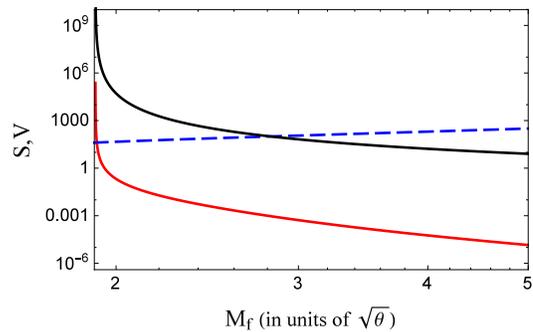}
\caption{(Color online) The
comparison between the noncommutative volume $V$ (black solid line) and
the associated entropy $S$ (red solid line).
The blue dashed line denotes the noncommutative BH entropy $S_{h}$ of Eq. (\ref{ncbe}).
}%
\label{fig4}%
\end{figure}

A refined description with noncommutative spacetime can incorporate
the second evaporation stage \cite{pn09}, which
offers a different scenario from the final explosion with a diverging
temperature as for a commutative
Schwarzschild black hole shown in Fig. \ref{fig1}.
Modification due to noncommutative spacetime kicks in
at the final evaporation stage, especially when $M_{f}$ approaches $M_{0}$.
In this regime, the temperature can be approximated as $T_{h}\simeq\alpha
(M_{f}-M_{0})$, with $\alpha=\frac{dT_{h}}{dM}|_{r_{h}=r_{0}}$.
The same analysis as in Ref. \cite{mkp07} gives for large $v$,
\begin{equation}
v\sim\frac{1}{\left(M_{f}-M_{0}\right)^{3}}.
\end{equation}
The final evaporation stage thus needs an infinite time,
although the net change to the black hole radius will only be several $\sqrt{\theta}$.
While counter-intuitive at first sight, this result is consistent with the third law of thermodynamics:
zero temperature state cannot be reached with a countable number of steps
or within a finite time. The statistical
entropy in Eq. (\ref{ncrve}) formally remains the same as in commutative spacetime,
except for the noncommutative modifications in the expression of the volume.
We thus immediately arrive at
\begin{equation}
V_{\mathrm{NCR}}\simeq3\sqrt{3}\pi\frac{M_{f}^{2}\left(  1-\frac{2r_{n}}%
{\sqrt{\pi\theta}}e^{-{\frac{r_{n}^{2}}{4\theta}}}\right)  }{\left(
M_{f}-M_{0}\right)^{3}}, \label{ncref}%
\end{equation}
for $v\rightarrow\infty$.
It constitutes an example for a
black hole with an infinite volume wrapped by a finite horizon
since $V_{\mathrm{NCR}}$ is divergent when $M_{f}\to M_{0}$,
as shown in Fig. \ref{fig4}.

From Eq. (\ref{ncrve}), we obtain
\begin{equation}
S_{\mathrm{NCR}}\sim\eta\alpha^{3}A_{h},
\end{equation}
with $\eta=\frac{\sqrt{3}\pi^{2}\left(1-\frac{2r_{n}}{\sqrt{\pi\theta}} e^{-{\frac{r_{n}^{2}}{4\theta}}}\right)}{15\times16\left(  1-\frac{2M}{\sqrt{\pi\theta}}e^{-{\frac{M^{2}}{\theta}}}\right)^{2}}
\simeq0.05$. It approaches infinite as well,
although at a slower rate as shown in Fig. \ref{fig4},
because  $\alpha\rightarrow\infty$ when $r_{h}\rightarrow r_{0}$
from $\frac {dM}{dr_{h}}|_{r_{h}=r_{0}}=0$ as discussed earlier.
The noncommutative entropy is larger than the
noncommutative BH entropy below a critical value near $M_{0}$
determined by the curve crossing.
A noncommutative black hole that evaporates down to smaller than
$M_{0}$ thus can house larger information storage capacity (than the BH entropy).
The crossing point shown in Fig. \ref{fig4} is approximate, since the graphed
volume associated entropy is estimated using (divergent) $v$ from the second evaporation stage only.

The divergent volume of Eq. (\ref{ncref}) is thus a general feature of noncommutative
black holes irrespective of their other details. This shows
noncommutative black hole can possess an infinite CR volume.
It belongs to a class of black holes with finite surface [BH
entropy Eq. (\ref{ncbe})] but an infinite interior.
Such a result supports the interpretation that the BH
entropy might be independent of the interior of a black hole.

\section{Conclusion and Discussion}

In conclusion we study the CR volume based on thermodynamics for Schwarzschild
black hole with noncommutative spacetime, which
allows for a well-described final evaporation stage and
results in a finite cold remnant at zero temperature.
We find the CR volume is similar in expression to the result
in commutative spacetime, except for a noncommutative parameter
dependent modification prefactor. The improved estimate for the
advanced time $v$ leads to a divergent volume.
Thus we show noncommutative Schwarzschild black hole represents
a class of black hole with an infinite volume wrapped inside a finite surface.

Our work also sheds light on the information loss paradox.
Assuming unitarity,
regardless of the \textquotedblleft firewall" \cite{amps13},
our result implies that information for a noncommutative black hole is stored
on the horizon, consistent with the proposed idea of
\textquotedblleft soft hairs\textquotedblright, and this information can be
taken away by Hawking radiations. Such a scenario has been studied
by several groups \cite{jdb93,zcy09,zcy11,gy13}, despite of the lacking
for a microscopic mechanism for information transfer.
If information were not taken away by radiations,
the assumed unitarity becomes questionable.

Irrespective of whether
radiation back reaction is included or not, we find
the interior of a commutative black hole houses insufficient number
of modes to account for the BH entropy as shown in Fig. \ref{fig2}.
Introducing noncommutative spacetime
but neglecting the important second evaporation stage does not change
such a conclusion as shown in Fig. \ref{fig3}.
The refined treatment for the final evaporation stage by using
noncommutative spacetime, on the other hand, results in a divergent volume,
which shows the finite BH
entropy is statistically independent of the black hole interior.
The remnant in a noncommutative black hole
needs an infinite time to reach, which
implies that although information might be stored in the interior of a black hole,
in order to preserve unitarity the complete process of evaporation needs an infinite time.
This seemingly unpleasant outcome essentially heralds the breakdown of unitarity
if information were housed in the black hole interior and not taken away by radiations.

\section{Acknowledge}

B.C.Z is supported by NSFC (No. 11374330 and
No. 91636213) and by the Fundamental Research Funds for the Central Universities, China University
of Geosciences (Wuhan) (No. CUG150630).
The work of L.Y. is supported by MOST 2013CB922004 of the National Key Basic Research
Program of China, and by NSFC (No. 11374176 and No. 91421305).

\end{document}